\documentclass{article}

\usepackage[top=2cm, bottom=2cm, left=1.5cm, right=1.5cm]{geometry}
\newcommand{\citet}[1]{\textcite{#1}}
\newcommand{\citep}[1]{\parencite{#1}}

\usepackage{amsmath,amsfonts,bm}

\def\eqref#1{equation~\ref{#1}}
\def\1{\bm{1}}

\DeclareMathAlphabet{\mathsfit}{\encodingdefault}{\sfdefault}{m}{sl}
\SetMathAlphabet{\mathsfit}{bold}{\encodingdefault}{\sfdefault}{bx}{n}

\DeclareMathOperator*{\argmax}{arg\,max}

\usepackage{hyperref}
\usepackage{cleveref}
\usepackage{url}
\usepackage{authblk}  % authors with affiliations
            
\usepackage[utf8]{inputenc} % allow utf-8 input
\usepackage[T1]{fontenc}    % use 8-bit T1 fonts
\usepackage{url}            % simple URL typesetting
\usepackage{booktabs}       % professional-quality tables
\usepackage{amsfonts}       % blackboard math symbols
\usepackage{nicefrac}       % compact symbols for 1/2, etc.
\usepackage{enumitem}
\usepackage{subcaption}
\usepackage{xcolor}
\usepackage{wasysym}

\usepackage[sorting=none,doi=false,isbn=false,giveninits=true,style=nature]{biblatex} 
\bibliography{refs.bib}

\usepackage{microtype}      % microtypography
\usepackage{graphicx}

\usepackage{amsmath,amssymb,amsthm,bm,dsfont}
\usepackage{array, booktabs, makecell}
\usepackage{multirow}

\usepackage{pifont}
\title{Towards decoding individual words from non-invasive brain recordings}
\title{Decoding the meaning of words from the non-invasive brain recordings of 723 participants}
\title{Scaling word decoding from brain activity: translating the EEG and MEG responses to 4 Million words in 723 participants}
\title{Scaling brain decoding: Translating non-invasive neural responses to 4.8M words in 723 participants}
\title{Decoding individual words from non-invasive brain recordings across 723 participants}

\author[1]{St\'ephane d'Ascoli}
\author[2,3]{Corentin Bel}
\author[1]{Jérémy Rapin}
\author[1]{Hubert Banville}
\author[1]{Yohann Benchetrit}
\author[2]{Christophe Pallier}
\author[1,3]{Jean-R\'emi King}
\affil[1]{Meta AI, Paris}
\affil[2]{CNRS, INSERM, CEA, Neurospin center, Gif-sur-Yvette}
\affil[3]{Laboratoire des Systèmes Perceptifs, École Normale Supérieure, PSL University}
\date{}

\begin{document}
\twocolumn
\maketitle

\begin{abstract}
% Seamlessly decoding language from brain activity could be transformative to patients who, after a brain lesion, lost their ability to communicate. 
% Yet, current brain-computer-interfaces are either invasive or suffer from limited decoding accuracy, real-time applicability and generalizability. 
% However, the limited scope of the corresponding studies makes decoding results difficult to evaluate.
% Deep learning has recently managed to decode language from the brain activity of a few participants. 
Deep learning has recently enabled the decoding of language from the neural activity of a few participants with electrodes implanted inside their brain.
However, reliably decoding words from non-invasive recordings %, irrespective of language, task and acquisition device 
remains an open challenge. 
% Deep learning has recently facilitated the decoding of language from brain activity. 
% However, decoding non-invasive neural recordings remains a major challenge. 
%Can such brain decoders effectively scale across individuals, languages, and recording techniques? 
%
% To address this question, we develop a new deep learning architecture to decode words from non-invasive brain recordings and evaluate it on 723 participants recorded with electroencephalography (EEG) or magnetoencephalography (MEG), as they read or listen to text in English, French, or Dutch.
%
% After curating these brain responses to 4.8M words, we train our model to decode their semantic features, as captured by a pretrained word embedding. 
To tackle this issue, we introduce a novel deep learning pipeline to decode individual words from non-invasive electro- (EEG) and magneto-encephalography (MEG) signals. We train and evaluate our approach on an unprecedentedly large number of participants (723) exposed to five million words either written or spoken in English, French or Dutch. 
%
% Our model achieves top-10 accuracies up to 60\% in the best participants out of a vocabulary of 250 words, outperforming existing methods by a wide margin, and enabling zero-shot generalization to unseen words. 
Our model outperforms existing methods consistently across participants, devices, languages, and tasks, and can decode words absent from the training set.
Our analyses highlight the importance of the recording device and experimental protocol: MEG and reading are easier to decode than EEG and listening, respectively, and it is preferable to collect a large amount of data per participant than to repeat stimuli across a large number of participants.
Furthermore, decoding performance consistently increases with the amount of (i) data used for training and (ii) data used for averaging during testing.
%
% This gain in decoding performance is consistent across recording devices, languages and tasks. 
%
Finally, single-word predictions show that our model effectively relies on word semantics but also captures syntactic and surface properties such as part-of-speech, word length and even individual letters, especially in the reading condition.  
Overall, our findings delineate the path and remaining challenges towards building non-invasive brain decoders for natural language. 
%Overall, our study delineates the path and remaining challenges to building non-invasive brain-to-text decoders.
    
\end{abstract}

\section{Introduction}

\begin{figure*}[htb]
    \centering
    \begin{subfigure}[b]{\linewidth}
        \includegraphics[width=\linewidth]{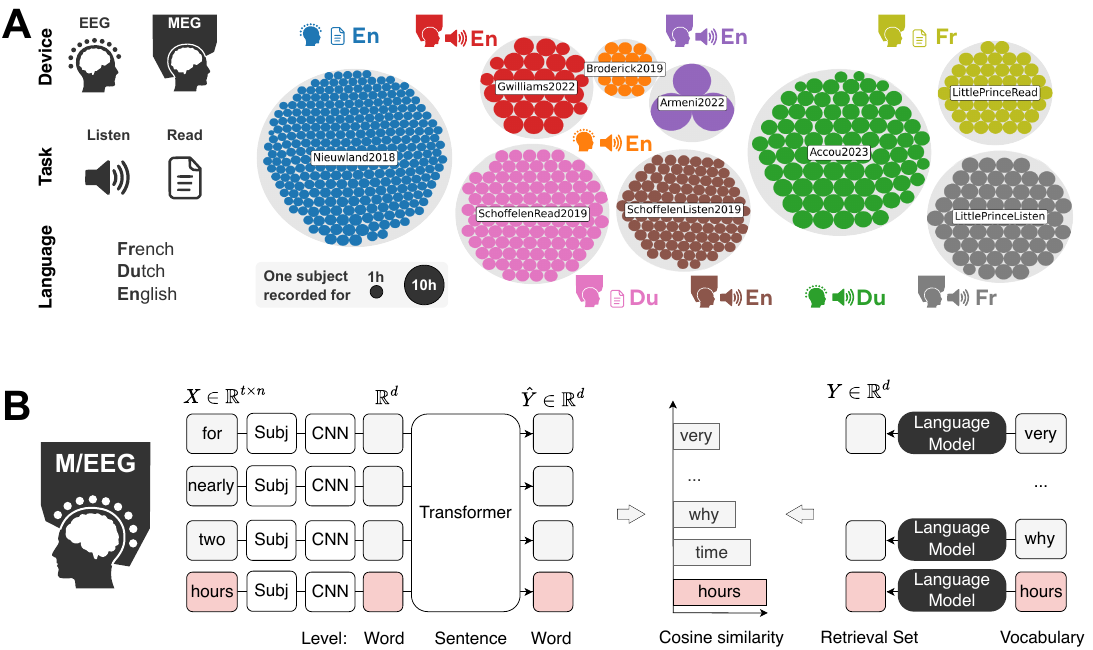}
    \end{subfigure}
    \centering
    \caption{
        \textbf{Approach.}\\
    \textbf{(A)} Each colored disk represents 1 subject (size represents recording time). Our datasets encompasses both public and original M/EEG data of participants reading or listening to Dutch, English or French sentences (Table \ref{tab:datasets}). \\
    \textbf{(B)} Our deep learning pipeline consists in training, with contrastive learning, an architecture that decodes the semantic representations of words from brain activity, as identified by a pretrained multilingual language model. \\
    % \textbf{(c)} For various sentences in the test set of LittlePrinceRead, we report the top-1 predictions on each word after aggregation over all subjects. For more examples, see \cref{app:predictions}.\\
    }
    \label{fig:sketch}
\end{figure*}

% Intracranial BCI work
In less than five years, artificial intelligence (AI) has been redefining the frontiers of brain-computer-interfaces (BCIs). 
Several groups have now demonstrated that intracranial implants in the motor cortex could be used to efficiently decode words from brain signals. For example, several machine learning algorithms can now learn to recognize patterns of neural activity associated with the intention to write or pronounce letters or syllables ~\cite{herff2019generating,angrick2019speech,anumanchipalli2019speech,metzger2023high,moses2021neuroprosthesis,willett2023high,wairagkar2024instantaneous,card2024accurate}. Such neuroprostheses could thus provide a voice to individuals who, after a brain lesion or a neurological disorder, have lost the ability to speak or communicate. 

% Intracranial is risky
However, such BCI requires an intracranial device, and thus neurosurgery. %Yet, opening the skull remains a risky procedure. 
In addition, cortical implants can be difficult to maintain beyond several months~\cite{chung2019high}. Non-invasive BCIs thus remain an important objective to assist or diagnose brain-lesioned patients \citep{owen2006detecting, king2013single, claassen2019detection, bodien2024cognitive}.

% Non invasive BCI are demanging 
Several non-invasive BCIs based on electro- or magneto-encephalography (M/EEG) exist but typically require users to perform tiring tasks over extended time periods, such as sustained visual attention or motor imagery~\cite{abiri2019comprehensive}. For example, the P3-Speller~\cite{marchetti2014effectiveness} is a popular protocol where participants can spell individual letters by paying attention to flickering stimuli on a computer screen. However, these approaches are too slow and too demanding to effectively scale to the constraints of natural language. 

% New approach: Word+-level decoding rather than spelling
This paradigm may be shifting, however. Over the past two years, two groups~\cite{tang2023semantic,defossez2023decoding} independently proposed a similar solution to directly decode sentences from non-invasive brain recordings of participants listening to natural language, by learning to align brain activations to those of AI language models. However, each of these two studies faced distinct limitations. 
%, the meaning of perceived and imagined words and sentences can, to some extent, be decoded from non-invasive brain recordings like functional Magnetic Resonance Imaging (fMRI) and MEG.

% Challenge of the new approaches: time limitation, and retrieval
\citet{tang2023semantic} relied on functional Magnetic Resonance Imaging (fMRI) to guide a language model during its text generation process. Although fMRI provides a good spatial resolution of the brain, its low temporal resolution stands as a barrier: each brain image is affected by all the words occurring in a multi-second time window, preventing the precise decoding of individual words. 

In contrast, \citet{defossez2023decoding} used EEG and MEG -- fast neuroimaging devices which have sufficient temporal resolution to access word-level details. However, the authors used a \emph{speech retrieval} strategy, where the decoder is trained to identify the most likely brain recording segment corresponding to a given speech segment. While this method achieves relatively good decoding performance, it requires access to the ground truth speech segments at test time, which limits its practical relevance. Additionally, it is unclear whether the decoder relies on the perceptual characteristics of the speech segment or the semantic features of the underlying words.

Finally, the scalability and robustness of language decoding remains unknown: most studies focus on a unique device, a single task, and a few participants. In sum, decoding, at scale, individual words from non-invasive approaches remains an open challenge.

To tackle this issue, we propose a new model optimized to decode individual words from EEG and MEG recordings. We validate our method with an unprecedentedly large dataset, encompassing 723 participants, either recorded with EEG or MEG while listening or reading sentences in their native language (Dutch, English or French).

\section{Methods}

\subsection{Problem statement}
The goal of the present study is to decode individual words from brain activity, including words absent from the training set. Such ``zero-shot'' decoding has already been shown to be statistically possible in~\cite{mitchell2008predicting,caucheteux2022brains,goldstein2022brain} using linear models, but the corresponding performances are too low to enable decoding in practice. 

Formally, we aim to predict pretrained word embeddings $Y \in \mathbb{R}^{d}$, from windows $X\in\mathbb{R}^{t\times n}$ of $t$ time-steps of brain activity recorded on $n$ sensors (note that this assumes knowledge of the word onset timings). 
Formally, we seek a mapping $f: \mathbb R^{t\times n} \to \mathbb R^d$ such that $\hat Y = f(X) \simeq Y$. This is illustrated in \cref{fig:sketch}B.

% We formulate the task of decoding individual words from 3-second windows of MEG as a contrastive learning problem, following the approach of~\cite{defossez2023decoding}. There are however two main methodological differences compared with \cite{defossez2023decoding}. 

%First, instead of using Wav2Vec embeddings of audio recordings as targets, we use the internal represenations of the raw words from a pretrained language model. The major differences this introduces is that the targets do not have any time dimension: our model is forced to extract semantics, rendering the task significantly more difficult, and requires some architectural changes.

% \subsection{Evaluations}

% Performance
% \begin{itemize}
%     \item top-10
%     \item balanced vs classic accuracy
%     \item single-trial vs aggregated
% \end{itemize}

% Interpretation
% \begin{itemize}
%     \item PCA
%     \item error decoding
%     \item Word frequency
%     \item zero-shot / out-of-vocab
%     \item cosine sim with word embedding
% \end{itemize}

% We introduce a new contrastive loss, which effectively handles the repetition of words and their extreme class imbalance -- a case which cannot be handled appriopriately by CLIP without discarding a significant amount of data. 

% Third, given the difficulty of our task, we evaluate on a dataset with a larger vocabulary, and introduce two new datasets to show the benefit of decoding from subjects reading instead of listening.

\begin{table*}[htb]
    \footnotesize
    \centering
    \begin{tabular}{ccccccccccc}
    \toprule
    Dataset & Subjects & \makecell{Time\\(h)} & \makecell{Words\\(k)} & \makecell{Unique\\words (k)} & \makecell{Unique\\sentences (k)} & Language & Task & \makecell{Narratives} & Device & Sensors \\
    \midrule
    \citet{Nieuwland2018Apr}                  & 295 & 171 & 706 & 0.9 & 0.3 & English   & Read     & False     & EEG & 63  \\
    \citet{broderick2018electrophysiological} & 19 & 20 & 214 & 1.6 & 0.8 & English     & Listen   & True    & EEG & 128  \\
    \citet{Accou2023Jul}                      & 80 & 150 & 1255 & 13.5 & 15.2 & Dutch   & Listen   & True    & EEG & 64  \\
    \midrule
    \citet{gwilliams2023introducing}          & 27 & 57 & 419 & 2.1 & 0.7 & English     & Listen   & True    & MEG & 208  \\
    \citet{armeni202210}                      & 3 & 34 & 247 & 7.3 & 5.0 & English      & Listen   & True    & MEG & 298  \\
    \citet{Schoffelen2019Apr}                 & 96 & 81 & 264 & 1.8 & 0.7 & Dutch       & Listen   & False    & MEG & 273  \\
    \citet{Schoffelen2019Apr}                 & 99 & 106 & 271 & 1.8 & 0.8 & Dutch      & Read     & False    & MEG & 273  \\
    LittlePrinceListen                        & 58 & 94 & 874 & 2.4 & 1.5 & French      & Listen   & True    & MEG & 306  \\
    LittlePrinceRead                          & 46 & 59 & 623 & 2.6 & 1.3 & French      & Read     & True    & MEG & 306  \\
    \midrule
    Total & 723 & 772 & 4877 & & & & & & \\
    \bottomrule
    \end{tabular}
    \caption{\textbf{Summary of the datasets considered.}}
    \label{tab:datasets}
\end{table*}

\begin{table*}
        \centering

        \begin{tabular}{p{0.48\textwidth}|p{0.48\textwidth}}
            \toprule
            \textbf{True words} & \textbf{Decoded words}\\ \toprule
            
            Yes it would be as well & \ \color[rgb]{0.00, 0.48, 0.82}{Yes}\ \color[rgb]{0.00, 0.48, 0.82}{it}\ \color[rgb]{0.08, 0.28, 0.44}{may}\ \color[rgb]{0.00, 0.48, 0.82}{be}\ \color[rgb]{0.39, 0.23, 0.08}{the}\ \color[rgb]{0.49, 0.27, 0.06}{guess}\ \\\hline
            Sherlock Holmes sat up with a whistle & \ \color[rgb]{0.00, 0.48, 0.82}{Sherlock}\ \color[rgb]{0.00, 0.48, 0.82}{Holmes}\ \color[rgb]{0.47, 0.26, 0.06}{stretched}\ \color[rgb]{0.00, 0.48, 0.82}{up}\ \color[rgb]{0.46, 0.25, 0.06}{on}\ \color[rgb]{0.21, 0.16, 0.12}{the}\ \color[rgb]{0.54, 0.29, 0.05}{wall}\ \\\hline
            Here we are said Holmes cheerily as we filed into the room & \ \color[rgb]{0.00, 0.48, 0.82}{Here}\ \color[rgb]{0.64, 0.33, 0.02}{her}\ \color[rgb]{0.71, 0.35, 0.01}{clock}\ \color[rgb]{0.00, 0.48, 0.82}{said}\ \color[rgb]{0.00, 0.48, 0.82}{Holmes}\ \color[rgb]{0.42, 0.24, 0.07}{rattling}\ \color[rgb]{0.74, 0.36, 0.00}{up}\ \color[rgb]{0.63, 0.32, 0.03}{the}\ \color[rgb]{0.74, 0.36, 0.00}{corner}\ \color[rgb]{0.00, 0.48, 0.82}{into}\ \color[rgb]{0.00, 0.48, 0.82}{the}\ \color[rgb]{0.14, 0.14, 0.15}{door}\ \\\hline
            He included us all in a sweeping bow and stalked out of the room & \ \color[rgb]{0.03, 0.41, 0.69}{She}\ \color[rgb]{0.29, 0.19, 0.10}{surveyed}\ \color[rgb]{0.74, 0.36, 0.00}{were}\ \color[rgb]{0.27, 0.18, 0.11}{sweeping}\ \color[rgb]{0.00, 0.48, 0.82}{in}\ \color[rgb]{0.00, 0.48, 0.82}{a}\ \color[rgb]{0.50, 0.27, 0.06}{hotel}\ \color[rgb]{0.37, 0.22, 0.09}{basket}\ \color[rgb]{0.00, 0.48, 0.82}{and}\ \color[rgb]{0.09, 0.27, 0.41}{limped}\ \color[rgb]{0.00, 0.48, 0.82}{out}\ \color[rgb]{0.00, 0.48, 0.82}{of}\ \color[rgb]{0.00, 0.48, 0.82}{the}\ \color[rgb]{0.53, 0.28, 0.05}{bridge}\ \\\hline
            He pushed past the servant and rushed into the drawing room followed by the king and myself & \ \color[rgb]{0.00, 0.48, 0.82}{He}\ \color[rgb]{0.13, 0.18, 0.23}{walked}\ \color[rgb]{0.57, 0.30, 0.04}{open}\ \color[rgb]{0.00, 0.48, 0.82}{the}\ \color[rgb]{0.61, 0.31, 0.03}{stair}\ \color[rgb]{0.00, 0.48, 0.82}{and}\ \color[rgb]{0.22, 0.16, 0.12}{brushed}\ \color[rgb]{0.00, 0.48, 0.82}{into}\ \color[rgb]{0.00, 0.48, 0.82}{the}\ \color[rgb]{0.36, 0.22, 0.09}{wooden}\ \color[rgb]{0.13, 0.17, 0.22}{stair}\ \color[rgb]{0.37, 0.22, 0.09}{pulling}\ \color[rgb]{0.41, 0.23, 0.08}{through}\ \color[rgb]{0.00, 0.48, 0.82}{the}\ \color[rgb]{0.46, 0.25, 0.06}{bridge}\ \color[rgb]{0.17, 0.14, 0.13}{of}\ \color[rgb]{0.49, 0.27, 0.06}{door}\ \\\hline
            I am ashamed of you Holmes said Lestrade with dignity after a few minutes' silence & \ \color[rgb]{0.21, 0.16, 0.12}{If}\ \color[rgb]{0.00, 0.48, 0.82}{am}\ \color[rgb]{0.67, 0.34, 0.02}{sign}\ \color[rgb]{0.00, 0.48, 0.82}{of}\ \color[rgb]{0.47, 0.26, 0.06}{mr}\ \color[rgb]{0.69, 0.35, 0.01}{office}\ \color[rgb]{0.00, 0.48, 0.82}{said}\ \color[rgb]{0.47, 0.26, 0.06}{remarked}\ \color[rgb]{0.67, 0.33, 0.02}{said}\ \color[rgb]{0.52, 0.28, 0.05}{ejected}\ \color[rgb]{0.00, 0.48, 0.82}{after}\ \color[rgb]{0.00, 0.48, 0.82}{a}\ \color[rgb]{0.00, 0.48, 0.82}{few}\ \color[rgb]{0.65, 0.33, 0.02}{marked}\ \color[rgb]{0.52, 0.28, 0.05}{resource}\ \\

            \bottomrule
        \end{tabular}
        % \caption{Examples of decoded sentences.}
        \label{tab:decoded_sentences}
    \caption{\textbf{Examples of decoded sentences.} 
    We report sentences from \citet{armeni202210}'s dataset, with the original stimulus on the left, and the decoded sentence on the right. 
    Colors of the latter denote semantic similarity measured by the cosine similarity between the true word and the predicted word, with red denoting semantically distant and blue denoting semantically close words.
    %Bold indicate incorrect predictions. Hue indicates the position in the sentence, and contrasts indicates the similarity, where blacks indicates no similarity between the true and predicted words. 
    }
\end{table*}

\subsection{Objective}

% \paragraph{Regression}

% A standard objective to achieve word decoding (\emph{e.g.}\cite{tang2023semantic}), is the a regression objective, i.e $\mathcal L_\text{MSE} =  \sum_{i=1}^N (Y_i-f(X_i))^2.$, with each batch of data consists of $N$ brain recordings $\{X_i\}_{i=1\ldots N}$ and $N$ word embeddings $\{Y_i\}_{i=1\ldots N}$. 

% However, \citet{defossez2023decoding} showed that this approach may be suboptimal because of its tendency to lead to regression-to-the-mean artefacts. Indeed, such losses are purely ``attractive'': to be more discriminative, the loss function needs a repulsion term too. 

% Following Defossez et al, we consider a CLIP (ref) objective:
% \begin{equation}
%     clip
% \end{equation}

% Given that the same word may be associated with different brain activity patterns, we consider SigLIP (ref):
% \begin{equation}
%     siglip
% \end{equation}

% Finally, to avoid informative discarding pair, we used D-SigLIP:
% \begin{equation}
%     dsglip
% \end{equation}

\paragraph{CLIP}

%Following \citet{defossez2023decoding}, we anticipate that this objective may be suboptimal, as it can lead to regression-to-the-mean artefacts. Indeed, such losses are purely ``attractive'': to be more discriminative, the loss function needs a repulsion term too. 
Following~\citet{defossez2023decoding}, we rely on contrastive learning to map the brain responses to the word embedding space. Defining the cosine similarity as follows:
\begin{equation}
    \measuredangle_{ij}= \frac{\hat Y_i\cdot Y_j}{\Vert \hat Y_j\Vert \Vert Y_j\Vert},
\label{eq:cosine}
\end{equation}

Our objective is to maximize the cosine similarity for positive pairs $i=j$ and minimize that of negative pairs $i\neq j$.

The CLIP loss~\cite{radford2021learning} treats this problem as a multiclass classification problem via the softmax function:
\begin{equation}
    \mathcal L_\text{CLIP} = -\frac{1}{N}\sum_{i = 1}^N \log\left(\frac{e^{t \measuredangle_{ii}}}{\sum_{j=1}^N e^{t \measuredangle_{ij}}}\right), 
\end{equation}
where $t$ is parameterized as $t=\exp(t')$ with $t'$ a learnable parameter.

\paragraph{D-SigLIP}
In our setting, there may be several repetitions of the same word in the batch, triggering different brain responses. This makes the contrastive learning problem ill-posed, as there would be matching (brain response and word embedding) pairs  with a negative label. 
%Removing repetitions would solve this problem, but would waste a lot of valid data in the process.

This issue can be addressed with a SigLIP loss~\cite{zhai2023sigmoid}. 
The SigLIP loss~\cite{zhai2023sigmoid} was originally introduced to improve the scalability of language-image pretraining to large batches. It treats each element of the batch as $N$ binary classification problems, which dispenses with the need to compute normalization factors across the batch:
\begin{equation}
\mathcal L_\text{SigLIP} = \frac{1}{N}\sum_{i = 1}^N \sum_{j = 1}^N \log\left(\frac{1}{1+ e^{z_{ij}(-t\measuredangle_{ij}+b)}}\right),
\end{equation}

where $b$ is a learnable bias and $z_{ij}$ is usually equal to $+1$ for $i=j$ and $-1$ for $i\neq j$. In our case, we can also use  $z_{ij}$ to define several positives for a given element of the batch. However, if we assign positive labels to all matching pairs, the fraction of positive to negative labels for a given word scales proportionally to the word frequency, which can cause class imbalance. 

To alleviate this issue, in case of repeated words, we discard the repetitions from the loss: we call this modified loss D-SigLIP, for ``Deduplicated SigLIP'', see \cref{fig:loss_ablation} for more details and ablations.

\subsection{Models}
\paragraph*{CNN}

To learn $f$, we first use the BrainModule model of~\cite{defossez2023decoding}\footnote{\url{https://github.com/facebookresearch/brainmagick}}. 
It consists of (i) a spatial attention module combining the data from the different sensors given their spatial positions, (ii) a subject-dependent layer which handles inter-subject variability, and (iii) a stack of convolutional blocks.

%First, we introduce positional embeddings which are added at the input layer to make the convolutional filters position aware. Indeed, the assumption of translation invariance which underlies convolutional layers breaks down here: the word to be decoded always starts exactly at the same time in the window, and accessing information on the timestep is crucial.
To transform the dynamical output of the CNN, of size $t\times d$, to a static word-embedding $\hat Y$ of size $d$, we need to pool the temporal dimension. For this, we use a single-head self-attention layer with unit output dimension. %More details on the learning model are reported in \cref{app:model}.

\paragraph*{Transformer}
The CNN only processes a single window containing the target word, and is not input with the surrounding context. To better exploit context, we propose a new decoding architecture which consists in adding a transformer on top of the CNN, as illustrated in \cref{fig:sketch}. Specifically, we split the text into sentences, each word of the sentence is fed to the CNN independently, and the outputs are stacked to form the input sequence of the transformer.

The transformer uses 16 layers and 16 attention heads, and its input dimension is equal to that of the target word embeddings, i.e. 1024. It uses an attention dropout of 0.1 and rotary positional embedding, following the implementation of the \texttt{x-transformer} package\footnote{\url{https://github.com/x-transformers}}. 

Note that this transformer is not a pretrained language model, and can thus be evaluated on its ability to retrieve information from brain signals, without leaking pre-existing linguistic knowledge.

\paragraph*{Baseline models}

We compare our deep learning pipeline to three baselines. 

The first is a standard ridge regression model, implemented via SKLearn's \texttt{RidgeCV}~\cite{pedregosa2011scikit}, using a logarithmic search for the regularization parameter \texttt{alpha} varying from $10^{-2}$ to $10^8$. It is used to predict each component of the target word embedding separately (with a different regularization parameter for each component) from a single slice of M/EEG data. We vary the offset of the timepoint relative to the word onset between $-0.5$ and $2.5$ seconds. 

The second is EEGNet~\cite{lawhern2018eegnet,schirrmeister2017deep}, a standard convolutional network used across the  brain decoding literature. Finally, we compare the results obtained with our full pipeline with those obtained using only the BrainModule~\cite{defossez2023decoding}, without the transformer on top.

\subsection{Evaluation}
\label{sec:metrics}

\paragraph{Word retrieval}
The models described above predict word embeddings, rather than words directly.  
To obtain the corresponding words, we select from a fixed vocabulary $\mathcal V$, by identifying the word whose embedding has the highest cosine similarity to the predicted embedding $\hat Y$: $\argmax_{Y\in \mathcal V} \measuredangle (\hat Y, Y)$

% When the model is fed with a brain recording $X$, we obtain the predicted word by searching for the word embedding $Y$ in a retrieval set $\mathcal R$ with the highest cosine similarity, i.e. $\argmax_{Y\in \mathcal R} \measuredangle (f(X), Y)$.

The above word-prediction step depends on the chosen vocabulary, as demonstrated in \cref{fig:retrieval_set_size}. However, vocabulary size varies substantially across datasets (see \cref{tab:datasets}). Consequently, when comparing datasets, and unless stated otherwise, we report our metrics at a fixed vocabulary size, by selecting for each dataset the 250 most frequent words. This reduced vocabulary covers between 70 and 95\% of all word occurrences for the various datasets considered. Words falling outside of the reduced vocabulary are discarded from the evaluation metrics.

%Note that this is significantly more than the 50-word retrieval set considered in recent work on invasive methods~\cite{moses2021neuroprosthesis}.

\paragraph{Accuracy}
Top-1 accuracy is typically too strict to assess semantic decoding, as it would count as incorrect situations where our model predicts a synonym of the true word. Hence, we use \textbf{top-10 accuracy}, which measures how often the correct word is within the model's top 10 predictions. 

Words in natural language are highly imbalanced: the ten most frequent words typically account for around 25\% of word occurrences.  Consequently, evaluation metrics may be biased by the most frequent words. 
%Indeed, consider a vocabulary of size $V$: since the ten most frequent words represent around 25\% of word occurrences in natural language, a trivial predictor which assigns a high score to the 10 most frequent words and zero to the others achieves a non-balanced top-10 accuracy near 25\% independently of $V$, while the balanced top-10 accuracy is equal to $10/V$, which is typically much smaller. 
Consequently, we compute a \textbf{balanced accuracy}, where accuracies are separately averaged per word, then averaged across words.

Unless specified otherwise, we will use this single-trial balanced top-10 accuracy on the 250 most frequent words throughout the paper, and will refer to it as ``Top-10 accuracy'', for read.

\paragraph*{Averaging over repetitions}
The above evaluations metrics focus on single-trial decoding, so as to provide a fair estimate of the decoding performance in a real-time setup. To further explore the extent to which our decoder capture a variety of linguistic features, such as semantics and part-of-speech, we also compute evaluation metrics of ``averaged-trials''. This approach improves signal-to-noise ratio by averaging the predictions of the model across the multiple brain responses to the same word. We either average repetitions across different contexts for a given subject 
(\emph{e.g.} subject 1: ``\textbf{To} be or not \textbf{to} be''), or across different subjects for a given context (Subject 1: ``\textbf{To} be or not to be''; Subject 2: ``\textbf{To} be or not to be''), or both.

\subsection{Datasets}

To evaluate our ability to systematically decode individual words from non-invasive brain recordings, we consider datasets that provide a large amount of temporally-resolved brain responses to language. For this, we surveyed the main public databases, namely Osf.io, Datadryad, OpenNeuro and the Radboud University Data Repository. We identified 7 relevant EEG or MEG datasets, where the task either involved speech comprehension or reading through a ``rapid serial visual presentation'' (RSVP) protocol, where words are flashed at the center of the screen, one after the other. We add to this two new MEG datasets where participants either listened to or read ``Le Petit Prince'', by Antoine de Saint Exupéry.  %This excludes free-reading datasets like Zuko \citep{zuko} where the gaze position may already inform the decoder about whether the brain activity corresponds to the end 

These datasets not only vary in terms of brain recordings (number of participants, number of recordings per participant, type of recording device), but also in terms of language stimuli (language, reading vs listening, decontextualized sentences vs narratives), see \cref{tab:datasets}. Consequently, these datasets allow us to evaluate decoding performances across a variety of data regimes, and experimental conditions.

All datasets were collected in accordance with participant consent and institutional ethics requirements. 
%

% We consider eight datasets, three of which are publicly available, and two new ones which we collected ourselves. Details on the datasets are provided in \cref{tab:datasets} and illustrated visually in \cref{fig:sketch}.
% [Where relevant, indicate the number of participants that were excluded because of difficulty in processing the data.]

\paragraph{Nieuwland}
% study summary
\citet{Nieuwland2018Apr}'s dataset is a preregistered EEG study involving nine distinct laboratories, which recorded a total of 334 participants\footnote{Due to preprocessing issues, we discarded the recordings coming from the University of Stirling, leaving 295 participants for our study.} while they read English sentences in an RSVP protocol. This study originally aimed to test whether the brain preactivate the phonological form of predictable words. 
% Ethics
%As indicated in the manuscript, "all participants were informed about the procedure of the experiment and then gave informed consent to use the data for research and dissemination/publication purpose. Ethical approval for EEG experimentation was obtained at each involved institution, according to custom guidelines of the ethics committee at each institution."

\paragraph{Broderick}
% Summary
\citet{broderick2018electrophysiological}'s dataset corresponds to the first EEG experiment of their study, where 19 subjects listened to the audiobook ``The Old Man and the Sea''. This study originally aimed to evaluate whether semantic features could be linearly retrieved from the EEG responses to words.
% Ethics
%As indicated in the manuscript, "all procedures were undertaken in accordance with the Declaration of Helsinki and were approved by the Ethics Committees of the School of Psychology at Trinity College Dublin, and the Health Sciences Faculty at Trinity College Dublin."

\paragraph{Accou}
\citet{Accou2023Jul}'s dataset, \emph{a.k.a} ``single-speaker stories dataset'' consists of 85 participants\footnote{The recordings of 5 of these participants are not publicly available.} who listened to audiobooks and podcasts in Dutch while being recorded with EEG. The goal of this study was to develop a deep neural network trained to decode the audio volume. Hence, the transcripts and word timestamps of the stimuli presented to the participants were not provided: we extract them using \texttt{WhisperX}~\cite{bain2023whisperx}.
% Ethics
%As indicated in the manuscript: ``Participants signed informed consent for this study, approved by the Medical Ethics Committee UZ KU Leuven/Research (KU Leuven, Belgium) with reference S57102. All data was collected and all experiments were performed in accordance with relevant guidelines and regulations.''

\paragraph{Armeni}
% Summary
\citet{armeni202210}'s dataset is a MEG study involving three participants who listened to ten one-hour long segments of ``The Adventures of Sherlock Holmes''. The goal of this study was to offer a dataset where a large amount of data is collected for each participant. 
% Ethics
%The manuscript indicates: ``in the written informed consent procedure, [the participants] explicitly consented for the anonymized collected data to be used for research purposes by other researchers. The study was approved by the local ethics committee (CMO — the local “Committee on Research Involving Human Subjects” in the Arnhem-Nijmegen region) and followed guidelines of the Helsinki declaration.''

\paragraph{Schoffelen}
\citet{Schoffelen2019Apr}'s dataset, \emph{a.k.a} ``Mother Of All Unification Studies'', consists of 204 participants who either read or listened to isolated sentences in an MEG scanner. This study was originally designed to explore the brain responses to a variety of syntactic and compositional structures. The participants who read the sentences are disjoint from those who listened to them, but the sentences are shared between these two subsets.
%"The duration of a single word (in ms) was determined as: (nletters/sumnletters) * (audiodur+2000-150 * nwords). No word was presented for a shorter time than 300ms. Each word was separated by an empty screen for 300ms before the onset of the next word."
% Ethics
%The manuscript indicates: ``In the informed consent procedure, [participants] explicitly consented for the anonymized collected data to be used for research purposes by other researchers. [...] The study was approved by the local ethics committee (CMO – the local “Committee on Research Involving Human Subjects” in the Arnhem-Nijmegen region) and followed guidelines of the Helsinki declaration.``

\paragraph{Gwilliams}
% Summary
\citet{gwilliams2023introducing} dataset, \emph{a.k.a} ``MEG-MASC'' consists of 27 participants who listened to approximately 2 hours of stories in the MEG scanner. Each story was repeated twice. The goal of this study was to provide a high-quality MEG dataset for encoding and decoding analyses. 
% Ethics
%As indicated in the manuscript: ``All participants provided a written informed consent and were compensated for their time. [...] The study was approved by the Institutional Review Board (IRB) ethics committee of New York University Abu Dhabi.''

% The three public EEG datasets feature recordings of participants reading decontextualized sentences~\cite{Nieuwland2018Apr}, listening to audiobooks~\cite{broderick2018electrophysiological} and podcasts~\cite{Accou2023Jul}. As for the four public MEG datasets, the first two feature participants listening to audiobooks~\cite{gwilliams2023introducing,armeni202210}, while the following two involve them reading or listening to decontextualized sentences~\cite{Schoffelen2019Apr}.

\paragraph{\emph{The Little Prince} datasets}
Using a 306 MEG Elekta Neuromag machine, we collected two MEG datasets in which 102 native French healthy volunteers were presented with the full story of \emph{Le Petit Prince} by Antoine de Saint-Exupéry  (see \cite{li_petit_2022} for a detailed description of the stimuli). 46 participants read the text, while 58 participants listened to it. 
%This text comprises 15.4k word tokens with a vocabulary size of 2.5k.

In the case of reading, a RSVP paradigm was used to avoid eye movements. Words were flashed at the center of the screen every 300\,ms  --- each word was displayed for 250\,ms followed by a 50\,ms blank screen  --- and sentences were separated by an additional blank screen lasting 500\,ms.  

Aside from the perceptual modality, the main differences between the two datasets are the following: (i) LittlePrinceListen has a larger total duration because it has more subjects but also because of the slower delivery of audio speech (100 min vs. 90 min for the visual presentation), (ii) the number of tokens per subject and vocabulary size differ because of the difference in segmentation of words\footnote{For example, "j'avais" is segmented as ["j'", "avais"] for the listening dataset and as ["j'avais"] in the reading dataset.}.

%Ethics paragraph.
These datasets were acquired at the Neurospin Center, CEA, Gif-sur-Yvette by authors CB and CP. The protocol (CEA 100 049 / ID RCB: 2018-A02586-49) was reviewed and approved by the Comité de Protection de Personnes Sud-Est VI Clermont-Ferrand (ethic committee). 

\paragraph{Preprocessing} The M/EEG recordings were first bandpass filtered between [0.1,40] Hz and resampled to 50 Hz, using built-in functions from $\texttt{MNE-Python}$~\citep{gramfort2014mne}, then scaled using \texttt{sklearn}'s \texttt{RobustScaler} and clamped in the range $[-5,5]$.

\paragraph{Splitting}
We split the recordings into 3\,s windows, where each window starts at the word onset, and baseline-correction is applied to the neural data over the first 0.5\,s. %In \cref{app:window}, we show how decoding performance varies with window onset and window duration, respectively.

We split the train, validation and test data with a 80/10/10 ratio. To avoid the data leakage observed in many language decoding papers~\cite{jo2024eeg}, we ensure that the same sentences presented to different subjects are assigned to the same split, by hashing them deterministically.

\subsection{Implementation details}

\paragraph*{Word embeddings}
We obtained word embeddings by processing individual (non-contextualized) words with HuggingFace's \texttt{t5-large} model~\cite{raffel2020exploring}, and extracting the hidden representations from the middle layer. This design choice allows us to (1) use a single model across languages and (2) deal with multi-token words. Indeed, when a word is split into several tokens, we average the contextualized embeddings of the resulting tokens to obtain the target word embedding. 
%We present ablations on the language model used in \cref{app:llm}, and noticed that causal models such as GPT2 work less well than non-causal models such as Bert and T5. While Bert provides less good results than T5, we did not find significant differences between the different sizes of T5 models, suggesting that the word embeddings do not benefit much from scale.

\paragraph{Training}
We train our model using the AdamW optimizer~\cite{loshchilov2017decoupled} using a learning rate of $10^{-4}$ and a batch size of $64$. We use a cosine learning rate decay over the first 50 epochs, and use early stopping based on the balanced top-10 accuracy on the validation set. 
%Visualizations of the training curves are provided in \cref{app:dynamics}. 
Each run presented in this paper takes a few hours on a A100 GPU with 40GB of RAM.

% \paragraph*{Baselines}
% We will compare this model with a linear baseline (similar to~\cite{tang2023semantic,goldstein2022brain}) and the popular EEGNet model~\cite{lawhern2018eegnet,schirrmeister2017deep} in ~\cref{fig:decoding}.

\begin{figure*}[h]
    \centering
    \includegraphics[width=\linewidth]{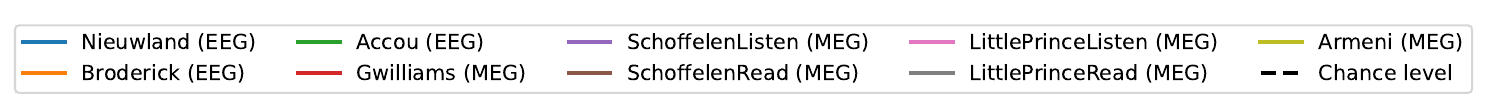}
    \includegraphics[width=\linewidth]{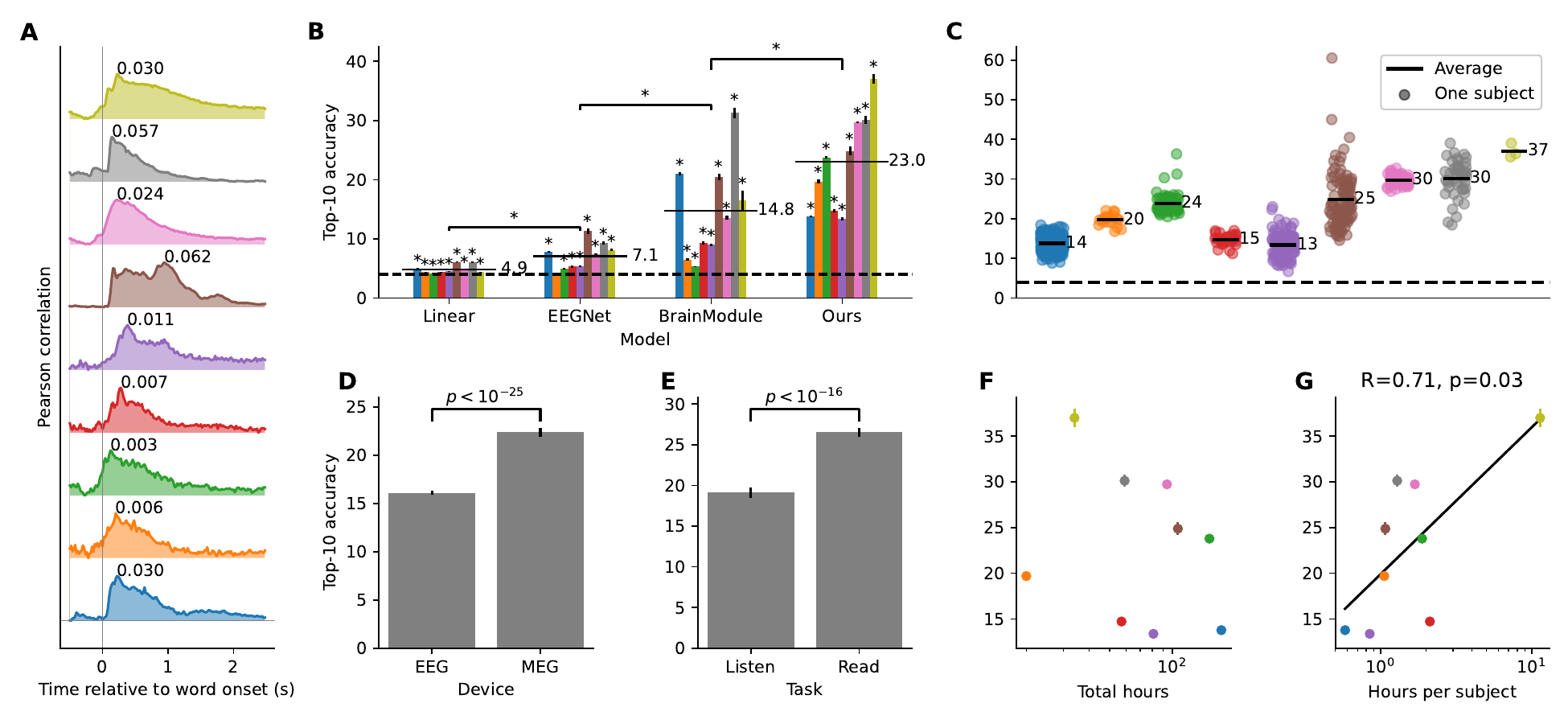}
    \caption{\textbf{Decoding performance across model architectures and datasets.} \\
    \textbf{A.} A linear ridge regression is trained to predict word embeddings from a single slice of M/EEG data, and at each time sample relative to word onset. Decoding is evaluated with the average Pearson correlation on the test set. We fit a different model for each subject, and report the average over all subjects. Each curve is normalized to its peak value, which is explicitly indicated above.\\
    \textbf{B.} We compare the accuracy for classic decoding models for each dataset (colors). Horizontal black lines denote the average across datasets for a given model. Stars highlight above chance decoding across participants ($p<0.005$).\\
    \textbf{C.} Accuracy of our model for each subject of each dataset, with the average over subjects denoted as horizontal lines.\\
    \textbf{D.} Accuracy averaged by recording device, with error bars denoting SEM across subjects.\\ % : MEG recordings are significantly better decoded than EEG recordings ($p<0.005$, Mann-Whitney U-test)
    \textbf{E.} Accuracy averaged by task, with error bars denoted SEM across subjects. We focus on MEG datasets that had the same sentences in the reading and listening condition. \\ % : reading tasks are significantly better decoded than listening tasks ($p<0.005$, Mann-Whitney U-test)
    \textbf{F.} Accuracy compared to the total recording duration of each dataset.\\
    \textbf{G.} Accuracy compared to the average recording duration per subject. The log-linear fit yields $p<0.05$.
    }
    \label{fig:decoding}
\end{figure*}

\begin{figure*}[h!]
    \centering
    \includegraphics[width=\linewidth]{final_figs/legend_horizontal.pdf}
    \includegraphics[width=.78\linewidth]{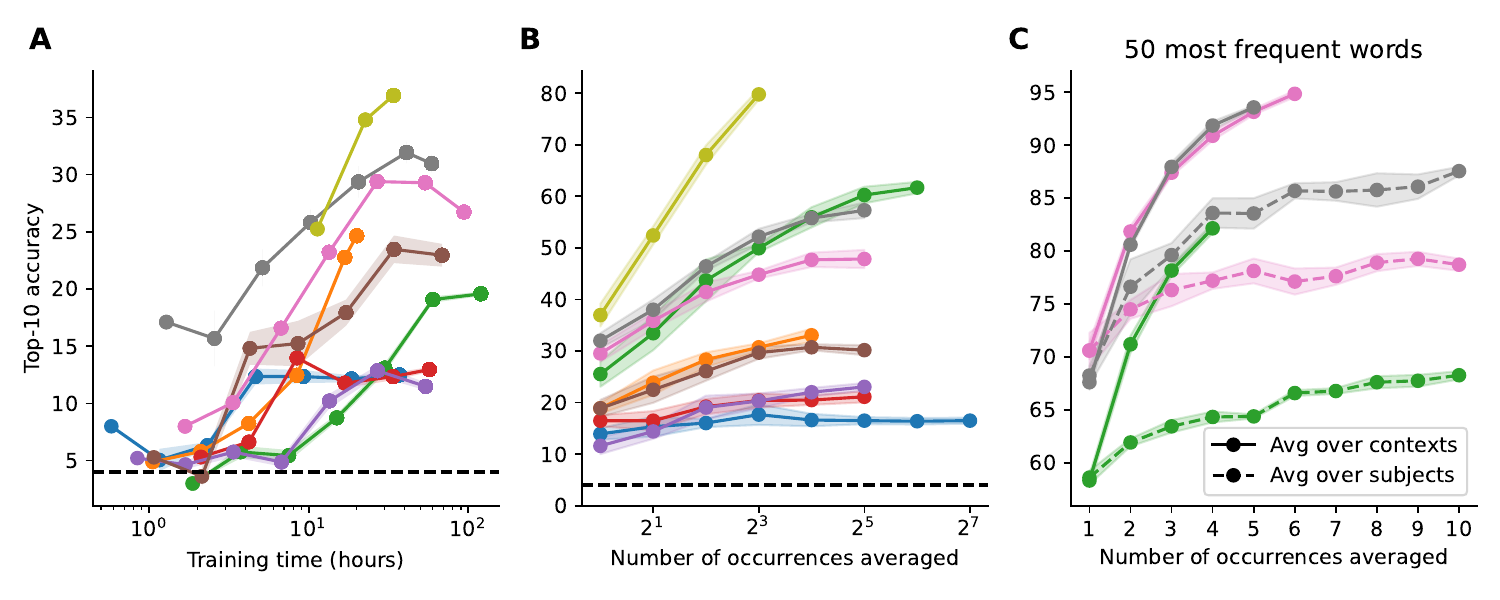}
    \includegraphics[width=.21\linewidth]{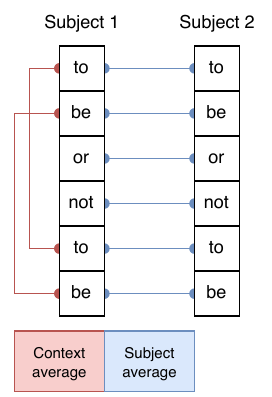} 
    \caption{\textbf{Scaling laws for decoding performance.} \\ % : Impact of data quantity and aggregation over word repetitions.
    \textbf{A.} Balanced top-10 accuracy as a function of the number of subjects used for training. Shaded regions indicate SEM across the subjects.\\
    \textbf{B.} 
    Balanced top-10 accuracy as a function of the number of occurrences of each word averaged before scoring. Shaded regions indicate standard deviation over the sampling of the occurrences.
    \\
    \textbf{C.} Comparison of averages within a given context and across $N$ subjects \emph{versus} averages within a given subject and across $N$  contexts for that subject, as illustrated by the sketch on the right. As this necessitates both the presence of many repetitions across subjects and contexts, we focus on the Accou and LittlePrince datasets which match these constraints, and consider the 50 most frequent words. Shaded regions indicate standard deviation over the sampling of the occurrences.}
    \label{fig:scaling}
\end{figure*}

\begin{figure*}[h!]
    \centering
    \includegraphics[width=\linewidth]{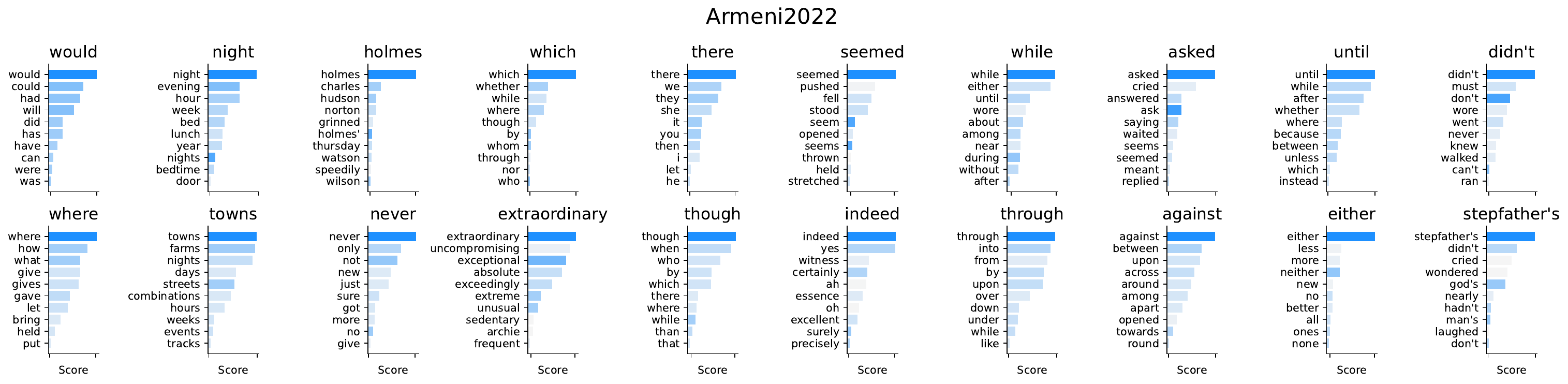}
    \includegraphics[width=\linewidth]{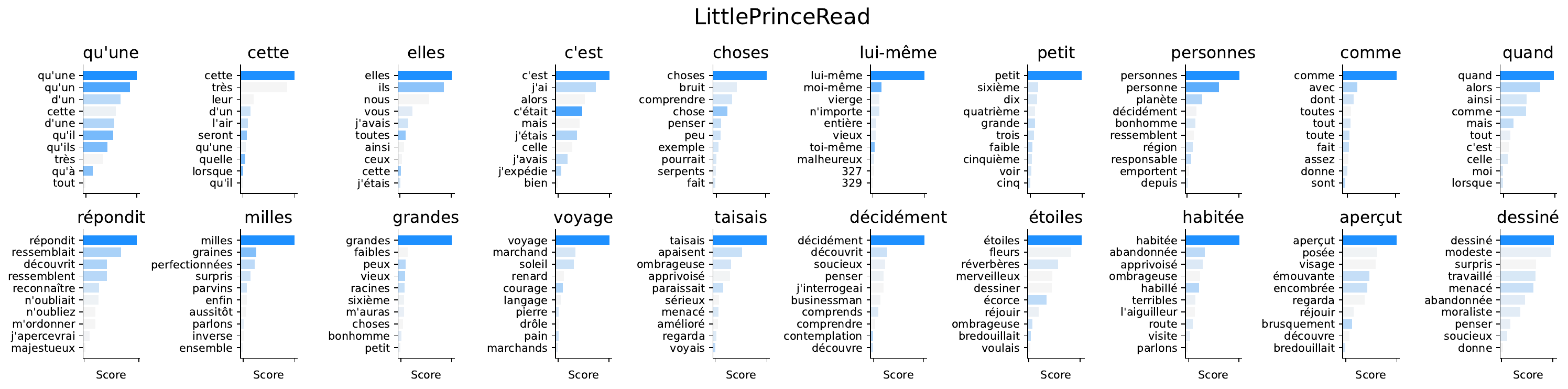}
    \includegraphics[width=.5\linewidth]{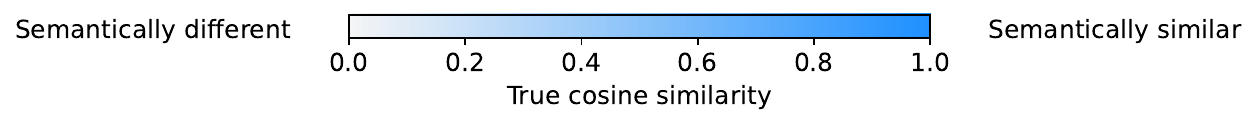}
    \caption{\textbf{Examples of top-10 predictions for two MEG datasets.} 
    The y-axis indicates the 10 most likely words given MEG activity. Horizontal bars represent the words $Y_j$ with highest cosine similarity to the decoding prediction $\hat{Y}_i$: $\mathbb E \left[ \measuredangle(\hat Y_i, Y_j) \right]$, where $\mathbb E$ denotes an average over all the recordings in the test set corresponding to word $i$. The colorscale indicates the true cosine similarity, i.e. $\measuredangle(Y_i, Y_j)$. Overall, many decoding predictions seem semantically similar to the words actually presented to the subjects. 
    More examples are displayed in \cref{fig:full_predictions}.
    }
    \label{fig:predictions}
\end{figure*}

\section{Results}

\subsection{Decoding performance across models and protocols}

\paragraph{Linear models}
We first aim to verify that word embeddings can be decoded from M/EEG signals with a linear model \citep{mitchell2008predicting,goldstein2024alignment,caucheteux2022deep,murphy2010detecting}. For this, we train a ridge regression model to predict the word embeddings from a unique time sample of the M/EEG recording at a time $\tau$ relative to the corresponding word onsets, and vary $\tau$ between $-0.5$\,s and $2.5$\,s. We evaluate the decoding performance by computing the Pearson correlation between each dimension of true and the predicted embeddings on the test set, then averaging across all dimensions. The resulting scores tend to peak within the first 500\,ms, although with varying performances across datasets (\cref{fig:decoding}A).
While ubiquitous in neuroscience, linear decoders may not be optimally designed to leverage the complex neural signals embedded in noisy M/EEG recordings. Indeed, a retrieval metric, based on the word embeddings linear models predict, leads to statistically significant ($p<0.005$) but extremely low top-10 accuracies: e.g. around 6\% for the best datasets, where chance level is 4\%  (\cref{fig:decoding}B). To address this issue, we now turn to the deep learning architectures.

\paragraph{Deep learning models}
We compare three deep learning pipelines, each trained with the same contrastive objective: EEGNet \citep{lawhern2018eegnet}, \citet{defossez2023decoding}'s ``BrainModule'', and our pipeline. All three models significantly outperform linear decoders (\cref{fig:decoding}B) ($p<0.005$, paired Wilcoxon test). \citet{defossez2023decoding}'s BrainModule, which is equipped with a subject layer, and thus designed to learn from different participants, significantly outperforms EEGNet ($p<0.005$), with a twofold increase in accuracy on average over datasets. Finally, our pipeline yields another 50\% performance boost on average over this BrainModule. This shows that our approach leads to a major improvement in comparison to existing M/EEG models. 

The decoding performance of our pipeline is well above chance for each of the 773 subjects included in our study (\cref{fig:decoding}C). Interestingly, the variability across subjects tends to be lower in the listening datasets than in the reading datasets: for example, for the LittlePrince datasets, accuracy varies between 26 and 33\% in the reading condition, and between 19 and 38\% in the reading condition. %We speculate that this may be related to the large volume of data per subject (10 hours).

Overall, these results highlight that our approach can reliably decode words from brain activity across a variety of participants, recording devices, tasks and languages. We provide a more extensive summary of metrics in \cref{fig:full_metrics}.

\paragraph{Impact of the experimental protocol}
How do the various aspects of the experimental protocol impact decoding performance?
Mann Whitney tests across participants show that our decoding pipeline achieves better decoding performances with MEG than EEG ($p<10^{-25}$, \cref{fig:decoding}D). Furthermore, it performs better when subjects read rather than listen to sentences (comparing datasets exposing subjects to the same sentences: $p<10^{-16}$, \cref{fig:decoding}E).
 
While decoding accuracy does not consistently vary with the datasets sizes,(\cref{fig:decoding}F), we observe a weak trend with the log volume of data \emph{per subject} (\cref{fig:decoding}G, $p<0.05$). This suggests that with a fixed recording budget, it is better to record a small number of participants across many sessions than a large number of participants with a small amount of sessions.

Together, these results highlight the impact of experimental designs such as recording device, task and time allocation per participant.

% For a complete summary of metrics, refer to \cref{app:metrics}, which also includes top-1 accuracies.

\subsection{Scaling laws of word decoding}
How does decoding performance scale with M/EEG data? To address this issue, we now focus on our decoding pipeline, and analyse how its decoding performance varies with 
(1) the amount of training data, 
(2) the amount of test words used to average decoding predictions, and
(3) the type of averaging used to improve decoding predictions.

\paragraph{Scaling training data}
We re-train a series of decoding pipelines on different subsets of our dataset, obtained by gradually increasing the number of subjects, starting with a single subject. The results show that decoding performance increases with the amount of training data (Fig.~\ref{fig:scaling}A), following a roughly log-linear trend. %This suggests that, in order to decode each participant's recording, our workflow efficiently extracts patterns of brain activity that are similar across participants. 
Although the trends vary across datasets, we do not observe clear signs of diminishing returns, hinting at the scalability of this decoding technique with the amount of experimental data.

\paragraph{Scaling test data averaging}
All of the decoding metrics reported so far correspond to \emph{single-trial} performances, such that they can be compared across datasets and inform real-time applications. 
However, studies often report decoding performances obtained from the \emph{average} of multiple identical trials (e.g. \citep{scotti2024reconstructing}). While such averaging cannot be directly translated to real-time conditions, it may clarify whether decoding failures are solely caused by the low signal-to-noise ratio of M/EEG (in machine learning terms, a high variance), or also reflect an imperfect learning of the mapping (a high bias).

In our setup, decoding performance steadily increases with the number of decoding predictions used for averaging, following a very clear log-linear trend. Most datasets show a two-fold improvement with such technique (\cref{fig:scaling}B). Remarkably, for the dataset of \citet{armeni202210}, the top-10 accuracy reaches close to 80\% after averaging only 8 predictions in response to the same word. These results suggest that decoding performance is strongly constrained by the low signal-to-noise ratio: reducing the latter via averaging improves performance drastically.

\paragraph{Averaging across subjects or contexts.}
In the previous analysis, we average the occurrences of words both over (i) repetitions for a given subject in response to different contexts and (ii) repetitions for several subjects in response to the same context (see \cref{fig:scaling}C for an illustration). How do these two averaging methods compare with each other?

To answer this question, we focus on the \citet{Accou2023Jul} and LittlePrince datasets, which contain both a large number of subjects and large amount of data per subject, and on the 50 most frequent words, which are repeated at least four times each for each subject.
Our results show that decoding performance increases substantially more rapidly when averaging over contexts than over subjects (\cref{fig:scaling}C). This explains why datasets which feature a large amount of sentences but few participants such as that of \citet{armeni202210} benefit more from averaging than those which feature a large amount of participants but few sentences such as that of \citet{Nieuwland2018Apr} (\cref{fig:scaling}B).

\begin{figure*}[h!]
    \centering
    \includegraphics[width=\linewidth]{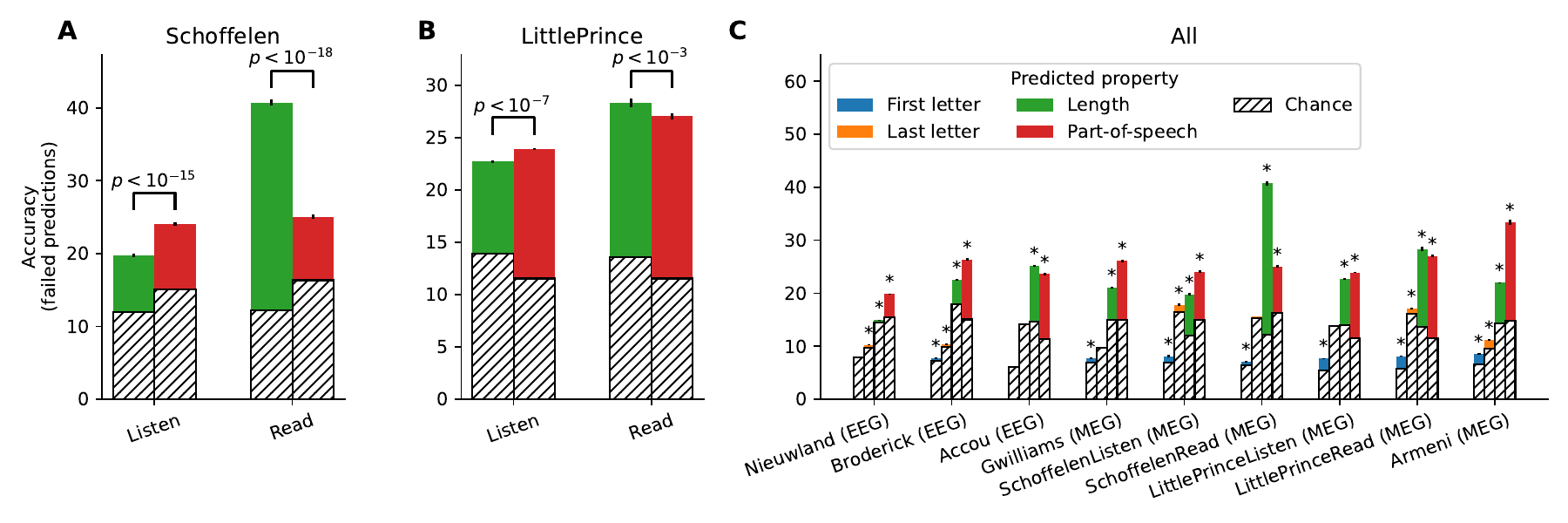}
    \caption{\textbf{Impact of sublexical and syntactic features on decoding.}\\
    For each incorrectly predicted word in the test set, we evaluate whether various properties of the top-1 prediction match those of the target word. \\
    \textbf{A-B.} Results for length and part-of-speech, where the stimuli are identical between the listening and reading tasks. Error bars denote SEM across subjects. %In both datasets, part-of-speech is better captured in the listening condition and length is better captured in the reading condition ($p<0.005$, paired Wilcoxon test). 
    \\
    \textbf{C.} Results for all properties (first and last letter, length and part-of-speech) across all datasets, with error bars denoting SEM across subjects.
    Stars indicate significantly above chance classification ($p<0.005$, one-sided t-test). 
    }
    \label{fig:mistakes}
\end{figure*}

\begin{figure*}[htb]
    \centering
    \includegraphics[width=\linewidth]{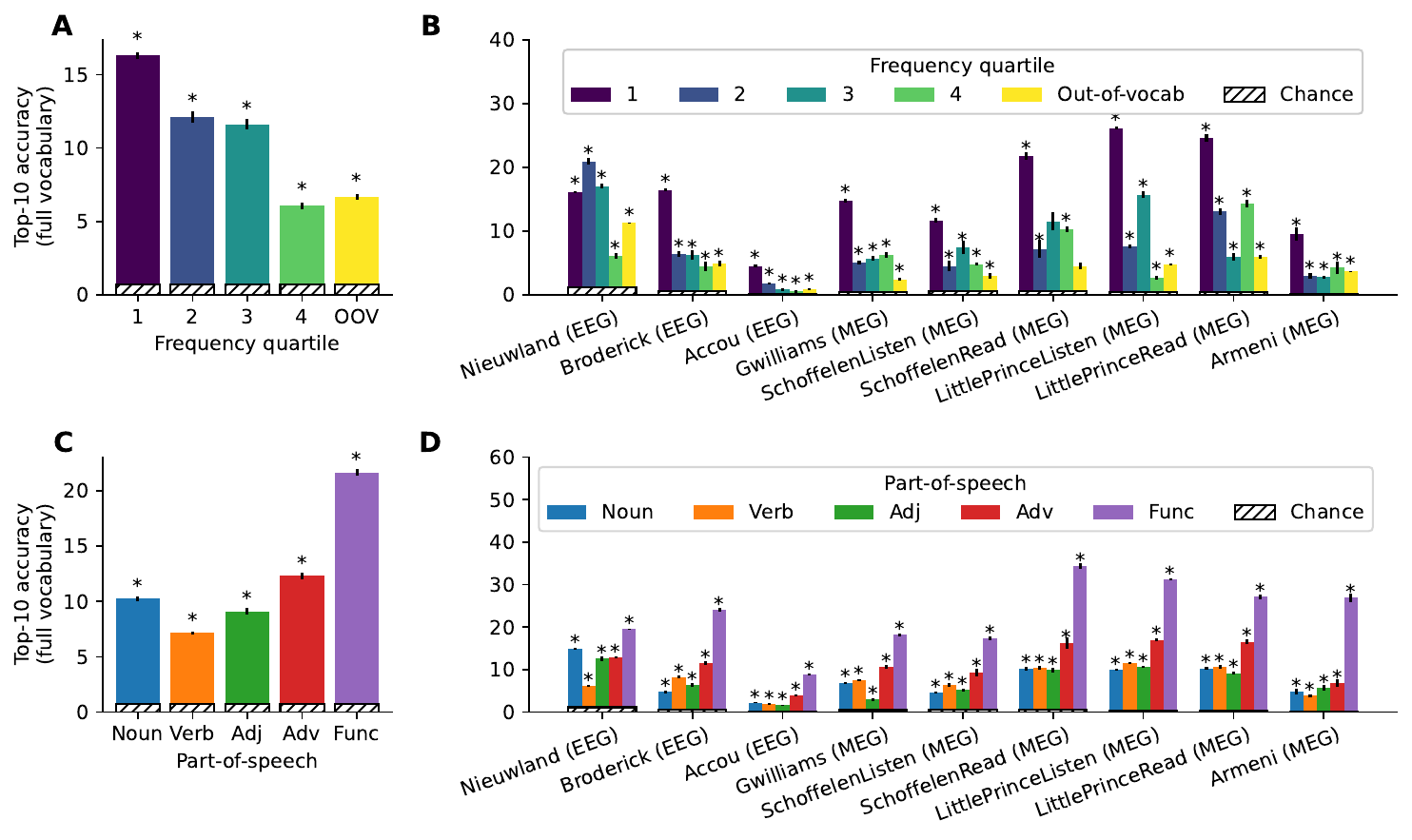}
    \caption{\textbf{Impact of various word properties on decoding performance.} \\
    \textbf{A-B.} Impact of number of occurrences of the word in the training set. We split the test set into quartiles according to the number of occurrences of each word in the training set, and denote as "out-of-vocabulary" words which do not occur in the training set. \\
    % \textbf{(b)} Impact of word length.\\
    \textbf{C-D.} Impact of the part-of-speech, obtained via Spacy. `Function' contains all part-of-speech types other than nouns, verbs, adjectives and adverbs. \\
    In all panels, error bars indicate the SEM over the subjects, and asterisks denote significantly better decoding than chance ($p<0.005$).
    }
    \label{fig:word_properties}
\end{figure*}

\subsection{Interpreting decoding performances}

Our transformer model can decode individual words from brain responses to natural language across a variety of experimental conditions. Do the decoders rely on word semantics?

\paragraph{Analysis of decoding predictions.}
To address this question, we first inspect the predictions of the decoder for individual words (\cref{fig:predictions}, more examples can be found in \cref{fig:full_predictions}). For simplicity, we focus this qualitative analysis on \citet{armeni202210} and LittlePrinceRead. 

For both datasets, many predictions appear to often fall close to the meaning of the word actually presented to the participants. For example, for the word `night', the top five predictions are `night', `evening', `hour', `week' and `bed'. Given that word embeddings are known to capture semantic features~\citep{mikolov2013distributed}, this result is coherent with the learning objective of our model.

For LittlePrinceRead, the predicted words appear to also capture visual features. For example, long words like `décidément' tend to be decoded as long but semantically-unrelated words. Similarly, for words containing hyphens, apostrophes and accents, the top predictions often contain these special characters.

% We observe that the model is able to capture semantics, as the highest ranked words are typically highly semantically related to the true word. 
% Note that our model is significantly more confident for the reading task, where the cosine similarity of the true word is up to 10 times higher than the average word, whereas for reading task it is only up to 25\% higher.

% In \cref{fig:rank_vs_similarity}, we demonstrate more quantitatively the relationship between the predicted rank and semantic proximity. We observe that words which have a cosine similarity above 0.8 are typically within the 10\% top predictions for \citet{armeni202210}. 

%Interestingly, while LittlePrinceRead displays the best single-trial decoding performance, \citet{armeni202210} achieves slightly better semantic consistency. This is elucidated in what follows: LittlePrinceRead leverages perceptual features to improve its decoding performance, which do not contain semantic information.

% \paragraph{Visualization}

% We provide a PCA visualization of the pretrained word embeddings as well as the embeddings predicted by our decoding model in \cref{fig:pca}. We see from the part-of-speech clusters than the overall structure of the embedding space is conserved by our model.

\paragraph{Analysis of mistakes.}

To quantitatively assess the relationships shared between the true words and their decoded predictions, we compare the incorrect top-1 predictions to their related true words (\cref{fig:mistakes}). We analyze four properties: whether the true and predicted words share the same first letter, the same last letter, whether they have the same number of letters, and finally, whether they share the same part-of-speech (\emph{i.e.} whether the words are both nouns, verbs, determinants, etc).

For almost all datasets, incorrect predictions share the same parts-of-speech and the same word length as the true word significantly more often than chance (\cref{fig:mistakes}C, $p<0.005$). This result suggests that incorrect predictions capture syntactic as well as some perceptual features. For some of the datasets, we also observe that sublexical features such as the first or last letter of the word are better decoded than chance. It is unclear whether this weak phenomenon is due to sensory representations, or to morphological properties (\emph{e.g.} in English, ending in 'ed' generally marks past tense and starting with 'wh' generally marks a question).

The comparison of the datasets presenting the same sentences to participants, either through a reading or a listening task, suggests that decoding reading relies more on visual features than decoding listening (\cref{fig:mistakes}A-B). Specifically, for both the LittlePrince and Schoffelen datasets, the length of the failed predictions matches that of the true words more often in the reading than in the listening condition ($p<0.001$), while the opposite holds when considering the part-of-speech ($p<10^{-7}$). Given the retinotopic structure of visual presentations, and the fixed font size and position, this result supports the idea that reading decoders also rely on visual responses.  

% The relevance of such perceptual features for decoding is reinforced by the models' attention maps, which focus on the M/EEG sensors close to the sensory cortices, namely the temporal channels for audio datasets and the occipital channels for the reading dataset (\cref{app:topomap}).

\paragraph{Relationship with word frequency.}
To test whether our model can decode words absent from the training set, we evaluate the \emph{zero-shot} decoding accuracy (\cref{fig:word_properties}A-B yellow). This analysis is designed to test whether we the decoder simply memorizes the exact M/EEG patterns elicited by each word, or whether it learns its underlying semantic features. Zero-shot decoding accuracy is significantly above chance ($p<0.005$), although its score is lower than for in-vocabulary words. This phenomenon appears to be linked to the fact that test words absent from the training set are typically rare in natural language, and hence harder to decode. Indeed, for in-vocabulary words, we observe that accuracy increases with the number of occurrences of the word in the training set (\cref{fig:word_properties}A). 

\paragraph{Relationship with part-of-speech}
Is decoding performance robust across different types of words? To address this question, we evaluate our decoder as a function of the part-of-speech categories (\cref{fig:word_properties}C-D) of the words considered. 
We observe a similar pattern for all datasets: performance is significantly above chance for all word lengths and part-of-speech categories, but is higher for function words ($p<0.005$). This result is consistent with the relationship between decoding performance and word frequency as function words are repeated many times in the training set. Overall, these results show that our decoder consistently decodes a variety of different words, but performs best with words repeated many times in the training set.

\section{Discussion}

In this study, we seek to evaluate the potential of a deep learning model to decode words from non-invasive recordings of a large number participants and across a variety of recording devices, tasks and languages. For this, we curated an unprecedentedly large M/EEG dataset, encompassing 723 participants reading or listening to isolated or contextualized sentences, and used it to train and evaluate a new architecture optimized to decode the meaning of individual words.

Our work provides two main contributions. Technically, it advances the development of BCIs, by showing that a single AI architecture can decode single words from brain activity recorded with a variety of non-invasive recording devices, and under a variety of experimental conditions. Theoretically, it offers insights to the neural underpinning of language representations.

\paragraph{Improving decoding performance from non-invasive recordings}
Technically, our decoding workflow outperforms classic methods such as linear models~\citep{goldstein2024alignment,caucheteux2022brains}, EEGNet~\citep{lawhern2018eegnet} and BrainModule~\cite{defossez2023decoding} by a large margin. This improvement is important to go beyond statistical metrics like the Pearson correlation, and develop evaluations applicable to BCIs. 

In this regard, our approach surpasses previous studies in two main ways critical to the development of real-time BCIs for natural language processing. First, and contrary to \citet{defossez2023decoding}, our approach is based on word semantics, and does not require having the true spoken sounds in a retrieval set. Second, and contrary to \citet{tang2023semantic}'s fMRI study and subsequent works~\cite{zhang2024decoding,ye2023language,chen2024open,yin2024language,xi2023unicorn}, our model aims to recover the semantics at the scale of \emph{individual} words, rather than sentences.

% Note that while some papers have attempted to perform sequence-level decoding of M/EEG~\cite{wang2022open,zhou2023belt,feng2023semantic,duan2023dewave}, these approaches are put into question as they typically suffer from improper train-test splitting~\citep{jo2024eeg}: the data they were trained on contains the same sentence repeated across subjects, which can simply be memorized by the decoder~\cite{varoquaux2017assessing,yin2024crosssubjectdatasplittingbraintotext}. 

Finally, our analyses highlight the impact of experimental designs. In particular, we show that MEG tends to be associated with better decoding performances than EEG. Similarly, the decoding of reading tends to be better than that of listening. Finally, decoding appears to be best when many recording hours are collected for each participant, rather than when many participants are recorded with the same stimuli for a short amount of time. 
%In particular, we observe a spectacular improvement when aggregating multiple instances of the same words across different contexts and subjects. This finding underscores the critical role of signal-to-noise ratio in decoding efficacy.

\paragraph{How AI helps modeling the neural bases of word semantics}
From a theoretical standpoint, our results confirms that the embeddings learnt by language models provide a remarkably useful tool to model semantic representations in the human brain~\cite{mahon_concepts_2008,binder_neurobiology_2011,binder2016toward,huth2016natural,ralph_neural_2017}. This approach was originated by \citet{mitchell2008predicting}, who showed that isolated words could be decoded from fMRI with a simple linear classifier targeting a word embedding trained with a latent semantic analysis. \citet{wehbe2014simultaneously, huth2016natural} observed that various types of word embeddings, learnt for example from co-occurrence statistics of words in natural language, can reliably account for brain responses to natural stories. Since, several studies systematically compared pretrained language models to the fMRI \cite{jain2018incorporating,caucheteux2021model,caucheteux2023evidence,millet2022toward,caucheteux2022deep}, intracranial \citep{goldstein2022brain,desbordes2023dimensionality,goldstein2024alignment} and M/EEG responses to spoken \cite{toneva2020modeling,caucheteux2022brains} and written sentences \cite{caucheteux2022brains}. Critically, several studies showed that this mapping between brain activity and word embeddings could be used to decode, in zero-shot fashion, words out-of-vocabulary \citep{mitchell2008predicting,caucheteux2022brains,goldstein2024alignment}.

Here, we further show that M/EEG responses to words lead to decoding predictions that can be surprisingly close semantically. %Given the relatively poor spatial resolution of MEG, these results strengthen the idea that words are represented as high-dimensional vectors embedded distributed across the cortical activations. 
The similarity between word embeddings and brain signals suggest that there exists general principles to the organization of semantic features in biological and artificial neural networks, and that these principles transcend the precise architecture or training schemes of these neural networks. The present study illustrates the exciting potential of A.I. in elucidating how the human brain structures symbols and knowledge.

Nevertheless, our decoding analyses further suggest that non-semantic features -- such as part-of-speech, word length and even single-letters -- also contribute to the decoders' predictions. These properties could both reflect sensory representations (\emph{e.g.} word frequency is correlated with the size of the word on retinotopic maps) or linguistic representations (\emph{e.g.} rare words may be more surprising). Consequently, how semantic representations are precisely articulated and disentangled from syntactic and sensory features remains an open challenge for future research.

\paragraph{Path and challenges for building non-invasive brain-to-text decoders}

Our decoding pipeline consistently scales with the amount of non-invasive brain recordings. However, there is a long way to go before this approach can be translated to practical applications. First, the present study focuses on language \emph{perception} and not on language \emph{production}. Second, single-trial performances remains far from those achieved with intracranial electrodes~\cite{ramsey2018decoding,makin2020machine,moses2021neuroprosthesis,tan2023decoding}.
For example, given a vocabulary of 50 words, we reach a top-1 decoding accuracy of 20\%, whereas \citet{moses2021neuroprosthesis} achieved a top-1 accuracy of 39.5\% with an electrode implanted in the motor cortex. 
The limited performance of a non-invasive decoder seem to be mainly challenged by signal-to-noise ratio. Indeed, averaging over multiple M/EEG responses to the same words rapidly lead to high accuracies (\cref{fig:scaling}B). 

%Nevertheless, our approach may present an advantage compared to invasive techniques: current invasive BCIs indeed rely on the decoding of sublexical representations (motor movements, phonemes) \citep{moses2021neuroprosthesis,metzger2023high,tan2023decoding}, whereas our approach focuses on decoding lexical representations, and could thus improve the amount of words one can hope to decode per minute. Confronting and perhaps combining the advantages of each of these approaches thus constitutes an important avenue for future research.

Several elements may partially improve the present decoding pipeline. First, we here focus on single-word decoding, and thus require knowing the timing of word presentations. Other approaches typically used in speech transcription, often use a a Connectionist Temporal Classification (CTC) loss to circumvent this issue. Second, while single-word decoding can be remarkable, the decoded sentences are often devoid of a clear meaning. This is expected, as a single word can suffice to break the grammatical structure of a sentence. To tackle this issue, language models input with the decoded predictions could improve the meaning of decoded sentences and narratives (\emph{e.g.} \citep{metzger2023high,tang2023semantic})

% \paragraph{Perspectives}
The biological bases of language, once an enigma, continue to yield its structure to the probing eyes of deep learning. This study indeed reveals the tantalizing prospect of decoding, at scale, the neural code of natural language, a feat that promises to not only democratize brain-computer-interfaces, but also expand our understanding of human cognition and its specificity in the animal kingdom.

% - Invasive
%     + communication
%         * Willett
%         * Moses
%         * Metzger
%         * other
%         -> limited to motor decoding
%     + semantics
%         * Goldstein
%         * zero-shot Nastase
% - Non-invasive
%     + communication
%         * Owen et al science 2008
%         * Claaseen et al NEJM 2019
%         * BCI review
%     + semantics

\section*{Acknowledgements}
The authors thank Fosca Al Roumi, Leila Azizi and Florent Meyniel for their support on the experimental protocol, Robin Schirrmeister and Loïc Barrault for useful discussions, and Pierre Louis Xech and Abhishek Charnalia for program management.

\clearpage

% \bibliographystyle{icml2024}
% \bibliography{refs}
\printbibliography

\clearpage
\onecolumn
\appendix
\section*{Supplementary information}
\begin{figure}[htb]
    \centering
    \includegraphics[width=\linewidth]{final_figs/legend_horizontal.pdf}
    \begin{subfigure}[b]{\linewidth}
        \centering
        \includegraphics[width=.4\linewidth]{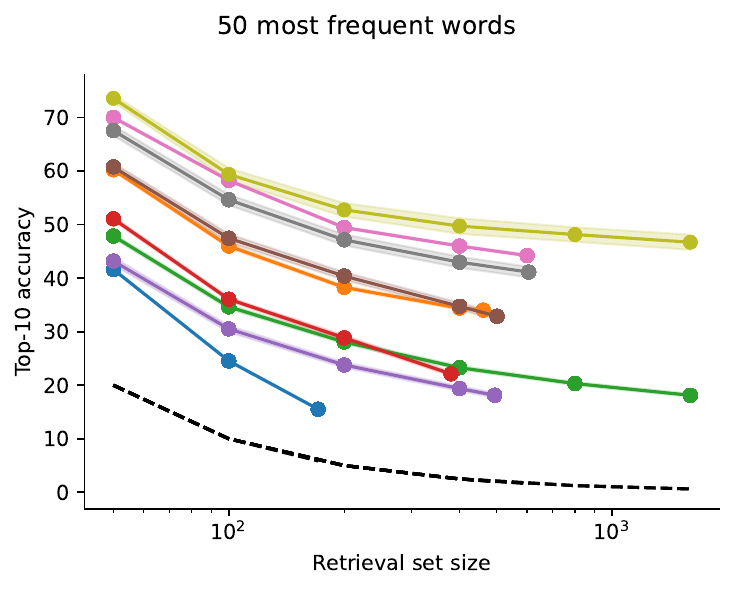}
    \end{subfigure}
    \caption{
    \textbf{Impact of retrieval set size.} \\
    Decoding performance on the 50 most frequent words as we increase the size of the retrieval set from 50 to 1600 words. Accuracy naturally decreases as the size of the retrieval set increases, but remains well above chance.
    % \textbf{(a)} We vary the duration of the window from $0.1s$ to $3s$, using a baseline of $0.1s$.\\
    % \textbf{(b)} We vary the offset of the window relative to the word onset, using windows of $1s$.\\
    % We gradually increase the number of subjects used for training and report test accuracy on the first subject.\\
    }
    \label{fig:retrieval_set_size}
\end{figure}

\begin{figure}[htb]
    \centering
    \begin{subfigure}{\linewidth}
    \centering
    \begin{tabular}{c|c}
    \toprule
        Loss &  Top-10 accuracy\\ \midrule
        CLIP & 15.7 $\pm$ 1.89 \\
        SigLIP & 14.7 $\pm$ 0.09\\
        D-SigLIP & 18.2 $\pm$ 0.76\\
    \bottomrule
    \end{tabular}
    \end{subfigure}
    \begin{subfigure}{\linewidth}
    \centering
    \includegraphics[width=.6\linewidth]{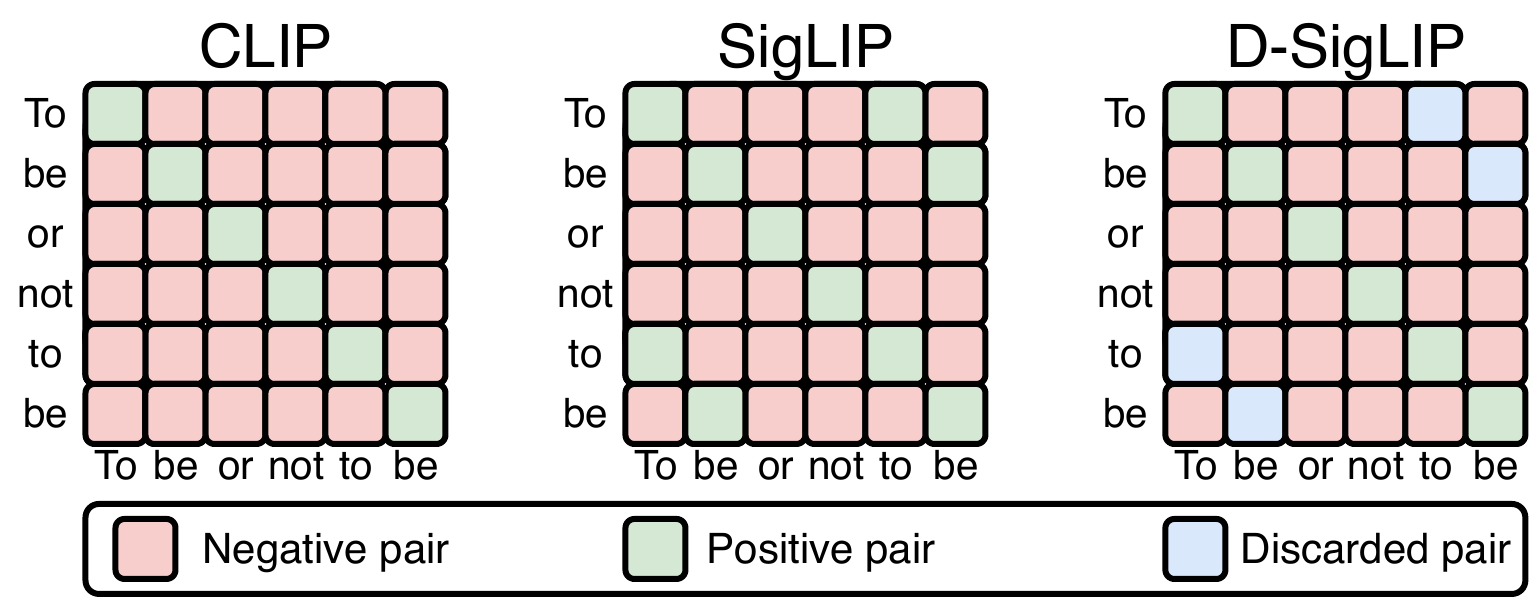}
    \end{subfigure}
    \caption{\textbf{Impact of the loss function.}\\
    We report the average top-10 accuracy on the most frequent 250 words across the participants of all datasets, as well as the SEM. Our D-SigLIP loss significantly outperforms the other ones ($p<0.005$).
    % The solid horizontal lines indicate average values over the datasets while the dashed horizontal line represent chance-level.}
    }
    \label{fig:loss_ablation}
\end{figure}

% \begin{table}[htb]
%     \centering
% \begin{tabular}{ccccc}
%     \toprule
%     Loss & \makecell{Multiple\\positives} & \makecell{Deduplicate\\positives} & \makecell{Non-balanced \\top-10} & \makecell{Balanced \\top-10} \\
%     \midrule
%     CLIP & \xmark & \xmark & 17.5 & 17.9 \\
%     SigLIP & \xmark & \xmark & 19.0 & \underline{18.0} \\
%     SigLIP & \cmark & \xmark & \textbf{23.8} & 13.5 \\
%     D-SigLIP & \cmark & \cmark & \underline{21.1} & \textbf{18.4} \\
%     \bottomrule
%     \end{tabular}
%     \caption{\textbf{Ablation on the loss function.} Results reported are computed for the most frequent 250 words from the Armeni2022 dataset.}
%     \label{tab:loss}
% \end{table}

% \section{Full metrics}
% \label{app:metrics}

% In Fig.~\ref{fig:metrics}, we report various metrics assessing the performance of our model on each of the datasets considered. 

\begin{figure*}[htb]
    \centering
    \includegraphics[width=\linewidth]{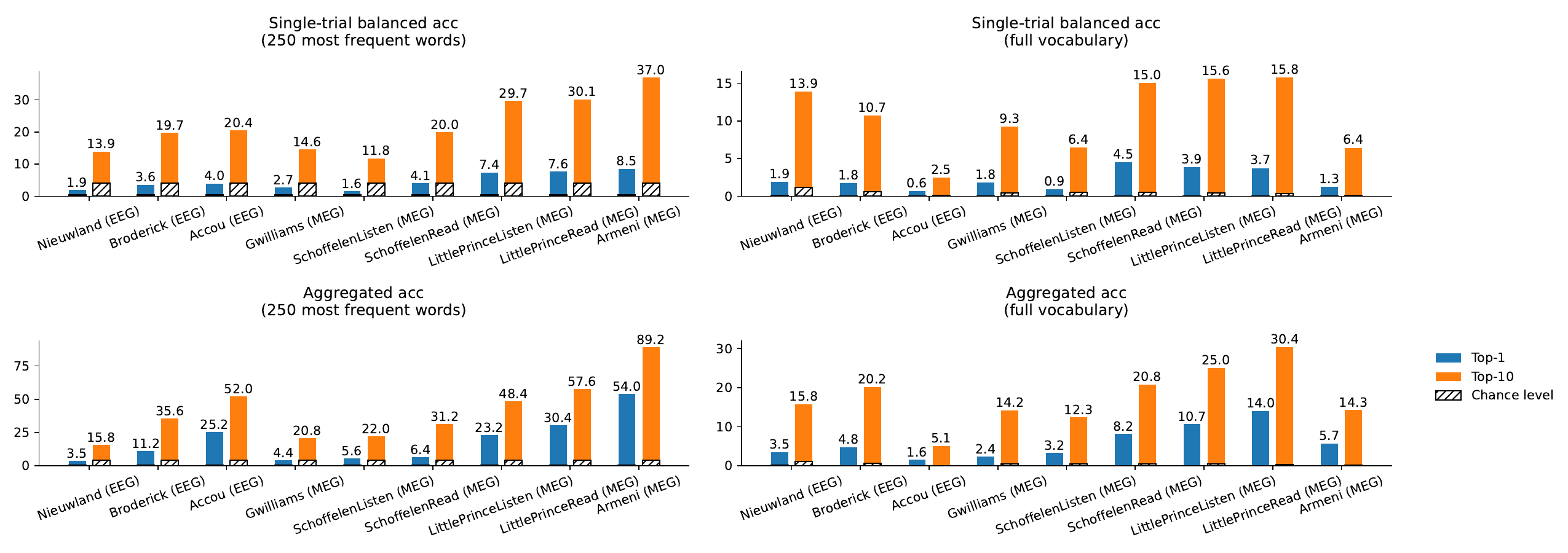}
    \caption{\textbf{Summary of decoding metrics for pipeline.} \\
    We report various single-trial and aggregated metrics as described in \cref{sec:metrics}, computed over the test set of each dataset.}
    \label{fig:full_metrics}
\end{figure*}

\clearpage

% \section{Learning model details}
% \label{app:model}

% We report the hyperparameters for the brain module in \cref{tab:brain_model_config} and those of the transformer in \cref{tab:transformer_config}. The overall architecture is the same as presented in~\cite{defossez2023decoding}.

% \begin{table}[htb]
%     \centering
%     \begin{tabular}{ll}
%     \toprule
%     Parameter    & Value \\
%     \midrule
%     activation & GeLU \\
%     depth & 5 \\
%     hidden size & 160 \\
%     dropout & 0.1 \\
%     % merger dropout & 0.2 \\
%     \bottomrule
%     \end{tabular}
%     \caption{\textbf{Hyperparameters for the brain module.}}
%     \label{tab:brain_model_config}
% \end{table}

% \begin{table}[htb]
%     \centering
%     \begin{tabular}{ll}
%     \toprule
%     Parameter    & Value \\
%     \midrule
%     depth & 16 \\
%     hidden size & 1024 \\
%     number of heads & 16 \\
%     attention dropout & 0.1 \\
%     % merger dropout & 0.2 \\
%     \bottomrule
%     \end{tabular}
%     \caption{\textbf{Hyperparameters for the transformer.}}
%     \label{tab:transformer_config}
% \end{table}

% \section{Predictions}
% \label{app:predictions}

%In \cref{fig:full-predictions}, we show more examples of top-scored predictions. 
% In \cref{fig:full-sentences}, we show how our model decodes words from individual sentences.

\begin{figure*}
    \includegraphics[width=\linewidth]{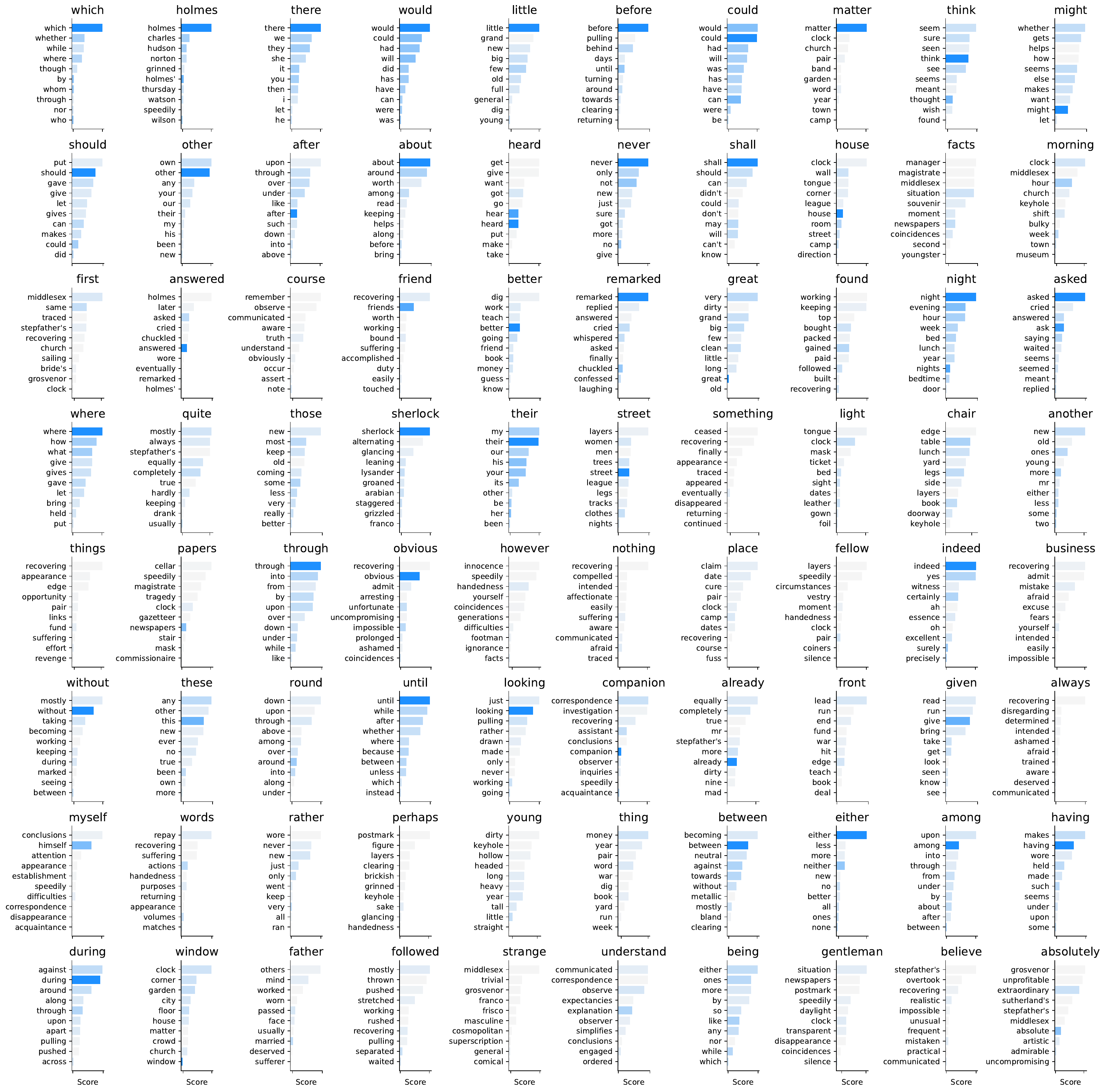}
    \caption{\textbf{Top predictions on a subset of words.}\\
    We selected the most 80 most frequent words from the test of \citet{armeni202210} containing at least five letters.
    }
    \label{fig:full_predictions}
\end{figure*}%

% \section{Understanding scores}
% \label{app:understanding}

% In our experimental protocol, we evaluated the understanding level of each subject using a set of questions. This gives us a score between 0 and 100 which we can subsequently compare with the decoding results achieved on each subject. We observe a positive linear correlation for LittlePrinceRead and no correlation for LittlePrinceListen as shown in \cref{fig:understanding}.

% \begin{figure}[htb]
%     \centering
%     \includegraphics[width=.8\linewidth]{figs/understanding.pdf}
%     \caption{\textbf{Correlation between the understanding score of the subjects and the decoding accuracy achieved.}}
%     \label{fig:understanding}
% \end{figure}a

\end{document}